# Examining the Effectiveness of Transformer-Based Smart Contract Vulnerability Scan

Emre Balci[a], Timucin Aydede[b], Gorkem Yilmaz[a] and Ece Gelal Soyak[a,*]

[a]*Bahcesehir University Computer Engineering, Istanbul, Turkey*
[b]*Bahcesehir University Artificial Intelligence Engineering, Istanbul, Turkey*

ABSTRACT

Smart contract technology facilitates self-executing agreements on the blockchain, eliminating dependency on an external trusted authority. However, smart contracts may expose vulnerabilities that can lead to financial losses and disruptions in decentralized applications. In this work, we evaluate deep learning-based approaches for vulnerability scanning of Ethereum smart contracts. We propose VASCOT, a Vulnerability Analyzer for Smart COntracts using Transformers, which performs sequential analysis of Ethereum Virtual Machine (EVM) bytecode and incorporates a sliding window mechanism to overcome input length constraints. To assess VASCOT's detection efficacy, we construct a dataset of 16,469 verified Ethereum contracts deployed in 2022, and annotate it using trace analysis with concrete validation to mitigate false positives. VASCOT's performance is then compared against a state-of-the-art LSTM-based vulnerability detection model on both our dataset and an older public dataset. Our findings highlight the strengths and limitations of each model, providing insights into their detection capabilities and generalizability.

## 1. Introduction

Smart contract technology is a fundamental innovation in blockchain, which enables self-executing contracts with the terms of agreement directly written into code. These decentralized programs eliminate the need for intermediaries in transactions, therefore enabling trustless and automated operations. Thanks to being deployed on blockchains such as Ethereum, these contracts are transparent and immutable, which facilitates a wide array of use cases across industries, including decentralized finance (DeFi), supply chains, Internet of Things (IoT) ecosystems, decentralized applications (dApps), and even non-fungible tokens (NFTs) that redefine the ownership of digital art.

Despite their benefits, smart contracts may sometimes contain errors, or the way they are invoked by numerous independent parties can trigger certain errors, rendering them vulnerable for exploitation by malicious entities. Such exploits may compromise the integrity of the contract's execution, leading to financial losses, data breaches, and disruptions to decentralized applications. Several incidents highlighted the impact of these vulnerabilities. To review examples, the most well-known vulnerability occurred in 2016 with the DAO hack, which resulted in a loss of approximately 60 million dollars worth of Ether due to a reentrancy vulnerability [1]. Similarly, in 2017, hackers exploiting the vulnerability in Parity Multisignature Wallet stole Ethereum tokens worth 30 million dollars [2]. The bZx protocol suffered from a flash loan attack in 2020, leading to substantial losses [3]. More recently, in Q1-Q2 2023, the DeFi industry lost about 735 million dollars due to exploits [4]. These incidents highlight the continuous threat of financial damage that can occur when attackers exploit vulnerabilities in smart contracts.

Furthermore, once smart contracts are deployed, it is not possible to update them (i.e., bug fixes cannot be deployed) due to the way blockchain works. Therefore, it is crucial to catch any vulnerabilities prior to deployment. Many research studies focused on this endeavour. Static and dynamic analysis techniques have been proposed to identify vulnerabilities such as reentrancy, integer overflow/underflow, and timestamp dependencies [5, 6]; these techniques may overlook certain vulnerabilities. Because some vulnerabilities arise only during specific sequences of contract invocations, symbolic analysis techniques have been proposed [7, 8]; these methods may be time consuming as the exploration of all executable paths is required. Among the various approaches to vulnerability scanning, machine learning-based sequential analysis of smart contract opcodes offers advantages in quickly detecting vulnerabilities that manifest during contract execution. Approaches that leverage Long Short-Term Memory (LSTM) networks have been used for learning patterns from large datasets of smart contracts [9].

In this work we perform an analysis of two deep learning based vulnerability scanning approaches. First, we propose VASCOT, a tool for the Vulnerability Analysis of Smart COntracts using the Transformer model. VASCOT performs sequential analysis of EVM bytecode. Next, to analyse VASCOT's detection performance, we collect recent contracts from the Ethereum network and label them using a vulnerability scanner that performs trace analysis. Trace analysis can output false positives, hence, we incorporate a complementary concrete validation phase. Finally, we evaluate and compare the performance of VASCOT with the state-of-art LSTM approach, on both our newly constructed dataset and an older public dataset.

The contributions of our work can be summarized as follows:

ORCID(s):

- We propose VASCOT, a vulnerability scanner that utilizes the Transformer model. As Transformer works with limited-size input sequences, we implement a sliding window mechanism to ensure VASCOT can properly analyze all input sequences.

- We construct a data set comprising 16,469 verified Ethereum smart contracts deployed in 2022. Having a recent dataset is important to capture the current state of contracts with the most recent language and compiler improvements.

- We compare the performance of VASCOT with that using LSTM on both our new data set, and on an older public data set. We discuss our observations, and highlight the challenges and the mechanisms that can be implemented to improve the generalizability of the vulnerability scanner models.

## 2. Related Work

Different techniques have been proposed for detecting vulnerabilities in smart contracts prior to deployment, to prevent incorrectly functioning smart contracts from being created. Earlier **static analysis** approaches examined the source code of a smart contract without executing it. Smartcheck [5] performed lexical and syntactical analysis on Solidity source code by translating it into an XML-based intermediate representation and checking against XML path patterns to identify vulnerabilities. ASGVulDetector [10] created an abstract semantic graph to capture the syntactic and semantic features of smart contracts. While it enables studying all possible execution paths, static analysis may not fully capture the interactions between contracts in runtime and thus, may miss vulnerabilities that occur under specific conditions. With **dynamic analysis**, functions are executed to identify vulnerabilities; however, execution has its own limitations, *e.g.*, when test cases do not account for all possible inputs and scenarios. As a result, several studies combined both approaches to improve the accuracy of detection. SolAnalyser [6] statically analyzed contract source code to assess locations prone to vulnerabilities, then dynamically executed all functions and transactions with different inputs. The framework proposed in [11] combined static and dynamic analysis to detect reentrancy vulnerabilities in Ethereum smart contracts. More recently, VULDEFF [12] proposed vulnerability detection based on function fingerprints and code differences.

Symbolic analysis is a type of static analysis that scans EVM *bytecode* to detect vulnerabilities. Oyente [7] used **symbolic execution** to capture traces that match the characteristics of defined vulnerability categories. Symbolic execution works by treating inputs and variables as symbols rather than actual values; it enables testing multiple execution paths at the same time rather than running the program on a specific concrete input. Mythril [13] and Manticore [14] are other examples of vulnerabilty analysis tools that used symbolic execution; the main drawback with these two tools was their time complexity. Zeus [15] used abstract interpretation and symbolic model checking to analyse Solidity contracts and operated in less time compared to its predecessors.

Many of these early symbolic analyzers modeled the behavior for a single invocation of a contract and were not designed to capture so-called **trace vulnerabilities**, *i.e.*, bugs that arise upon a sequence of invocations of a contract. To address this open question, Securify [16] was designed to decompile EVM bytecode into a stack-less dependency graph representation and scan this graph for violation patterns. MAIAN [8] proposed inter-procedural symbolic analysis on EVM bytecode, to find contracts that violate specific properties of traces. The tool flagged vulnerable contracts with three categories; the contracts that lock funds indefinitely (*greedy*), contracts that leak funds to arbitrary users (*prodigal*), and contracts that can be killed by any user (*suicidal*). The shortcoming of this approach is that, in practice there can be a large number of parallel invocations of a contract's public functions, and it may not be possible to model all possible traces for analysis in limited time. To maintain low levels of analysis duration, the tool has to maintain low invocation depth. To address this limitation, AI-based techniques have been proposed.

The use of **deep learning** (DL) for detecting vulnerabilities was first proposed in [9], inspired by the studies that perform DL-based malware detection. The study proposed sequence learning on EVM opcode using the Long Short Term Memory (LSTM) model. The model trained and tested on a public data set yielded 99.57% accuracy and F1 score of 86.04%. In our work, we show that these results are not viable on a *verified* data set, and that VASCOT outperforms LSTM in sequential analysis. Other studies also investigated sequential analysis of smart contract code using deep learning. BLSTM-ATT [17] transformed EVM source code to snippet representations capturing semantic information and control flow dependencies, then applied bidirectional LSTM with attention mechanism. AWD-LSTM [18] used Average Stochastic Gradient Descent Weight-Dropped LSTM and considered only distinct opcode combinations for normal contracts to reduce the class imbalance.

Machine learning and deep learning were also used for smart contract vulnerability detection in approaches other than the sequential analysis. ContractWard [19] extracted bigram features from opcodes, applied SMOTE and undersampling, and compared different supervised learning models on this data. This approach had limited efficiency due to requiring initial processing, as features would need to be re-evaluated as contract languages and capabilities evolve. A different approach was proposed in [20], studying smart contract vulnerability detection as an image classification problem. A multi-objective detection neural network (MODNN) architecture converted the operation sequence into explicit features, and constructed co-occurrence matrices using opcodes for implicit features. Some other studies highlighted the importance of fully considering the syntactic and semantic information of the program rather than opcodes. Blass [21] framework constructed program slices with complete semantic structure information and utilized an attention mechanism to capture the key features of vulnerabilities. Similarly, ESCORT [22] proposed extracting features that capture the semantics of vulnerability classes and trained a multi-output neural network architecture for detecting these types; this work also proposed transfer learning for scaling to new vulnerability types. EA-RGCN [23] also constructed a semantic graph for each function and trained Residual Graph Convolutional Network using an edge attention module. These tools only work on source code and consider a limited set of vulnerabilities.

The only study that used transformers in the scope of smart contract vulnerability detection was ASSBert [24], which tackled a different problem, the scarcity of labeled smart contract data for real-world vulnerability classification tasks. The proposed method utilized active learning for manual labeling of data, and semi-supervised bidirectional encoder representation from transformers network to select valuable code to add to training set.

## 3. Smart Contract Vulnerabilities

We provide a brief overview of how smart contracts are executed on the Ethereum Virtual Machine (EVM) environment, and how their compiled code may contain patterns that may indicate vulnerabilities.

### 3.1. Smart Contracts and the Ethereum Virtual Machine (EVM)

Smart contracts are stored on a blockchain, and are executed automatically and in a decentralized manner when predetermined terms and conditions are met. Ethereum Virtual Machine (EVM) is the runtime environment for executing smart contracts on one of the largest and most influential blockchains, the Ethereum network. Smart contracts on the Ethereum network are typically written in Solidity or Vyper, and are compiled to bytecode that can be executed on the EVM. When a smart contract is deployed, its bytecode is stored in a specific address; users interact with smart contracts by sending transactions (*i.e.,* data or Ether) to the contract's address.

```solidity
// SPDX-License-Identifier: MIT
pragma solidity ^0.8.0;
contract HelloWorld {
    string message;
    constructor(string memory initMessage) {
        message = initMessage;
    }
    function setMessage (string memory newMessage) public {
        message = newMessage;
    }
    function getMessage () public view returns (string memory) {
        return message;
    }
}
```

Algorithm 1: Source code of Hello World contract.

Algorithm 1 shows a contract named *HelloWorld* written in Solidity[1]. The contract contains a single string variable called *message* that is initialized with a value passed to the constructor function. The *setMessage* function allows external parties to set the value of the message variable, and *getMessage* allows external parties to read its current value. The Solidity compiler *solc* converts each function and variable in the source code to its corresponding bytecode representation according to the rules of the EVM. After compilation, the source code may or may not be publicly available on Etherscan, but the EVM bytecode is stored on-chain.

---
[1]In this work, without loss of generality, we focus on the smart contracts written in Solidity and executed on EVM.

EVM bytecode is a low-level representation of the program, comprising numeric codes, constants, address references that are generated as the compiler parses and performs a semantic analysis of type, scope, and nesting depths of program objects. Figure 1 represents part of the bytecode of the Hello World smart contract in Algorithm 1.

```
608060405234801562000011576000080fd5b5060405162000bf23803
```

**Figure 1:** Part of the Hello World contract converted to EVM bytecode.

The instructions of the hexadecimal bytecode representation are often mapped into a human readable form which is referred to as *opcode* [25]. Each byte in the bytecode represents a different EVM instruction (opcode) from the Ethereum Yellow Paper [26], for operations such as arithmetic (*e.g.,* ADD, SUB, MOD), stack/memory access (*e.g.,* SLOAD, MLOAD, MSTORE), control flow (*e.g.,* JUMP, STOP, SELFDESTRUCT), or other tasks (*e.g.,* ORIGIN, CALLER, GASLIMIT). Each opcode has a unique value; for example, the opcode for pushing a value onto the stack is PUSH1, with hexadecimal representation of 0x60. When the EVM encounters the 0x60 opcode, it reads the next byte as the value to be pushed onto the stack. Similarly, when the EVM encounters the hexadecimal 0x52, representing the MSTORE opcode, it writes the unsigned integer in memory. The bytecode 6080604052 (From Figure 1) represents the opcode sequence "PUSH1 0x80 PUSH1 0x40 MSTORE", i.e., PUSH the decimal value corresponding to the hexadecimal 0x80 (*i.e.,* 128) onto the stack, then PUSH the decimal value represented by 0x40 onto the stack, then allocate 128 bytes of memory space and move the pointer to the beginning of the $64^{th}$ byte. EVM documentation [27] and the Ethereum yellow paper [26] present the list of EVM opcodes and their respective hexadecimal values.

### 3.2. Smart Contract Trace Vulnerabilities

Smart contract vulnerabilities are flaws in their code that can compromise the security of the contract, render them susceptible to exploitation, potentially leading to financial loss. A list of smart contract vulnerabilities is provided in [28]. These bugs or vulnerabilities can be categorized in terms of how they are generated [29], following the IEEE Standard Classification for Software Anomalies [30].

Smart contracts can invoke other smart contracts through calls or other message-passing mechanisms. This invocation generally involves multiple transactions, which potentially modify the state of the contract. This system functions similarly to the concept of shared-memory concurrency in computers, and the errors commonly seen in concurrent programming, such as atomicity, synchronization problems, or resource ownership conflicts, are also seen in the transactions on EVM [31].

Consider a smart contract containing the function in Algorithm 2 that allows users to transfer tokens between accounts. Data flow analysis (e.g., tracking the use of *_to* and *_value* values) would help to capture how values flow through the code, e.g. the order in which the statements are executed, or whether certain statements can be executed conditionally based on the values of certain variables.

```
function transfer(address \_to, uint256 \_value) public returns (bool) {
    require(_value <= balances[msg.sender]);
    balances[msg.sender] -= _value;
    balances[_to] += _value;
    emit Transfer(msg.sender, _to, _value);
    return true;
}
```

Algorithm 2: Function allowing transfer of tokens between funds.

A *trace* is a sequence of invocations of a contract recorded on the blockchain. Trace vulnerabilities in contract source code refer to weaknesses that can be exploited by observing *i.e., tracing* the sequence of operations and the respective state of the contract on the blockchain, during the execution of a smart contract. Such trace-based attacks can lead to vulnerabilities like *front-running*, where attackers anticipate and preempt valid transactions based on their traces, or *reentrancy* attacks that exploit the sequence of function calls and states.

According to their representative characteristics, *trace vulnerabilities* have been categorized into three main groups [8]; those that permanently lock cash, those that leak funds to random users, and those that have the potential to be terminated by anyone. In this work, we focus on these categories of trace vulnerabilities that are explained next.

**Greedy smart contracts** are contract types where the party to the contract can no longer send funds. An example is the *Race Condition* vulnerability, where two or more actions can be executed concurrently or in an unpredictable order, leading to unexpected or undesirable outcomes. To prevent race condition vulnerabilities, locking mechanisms and atomic operations must be used to ensure concurrent operations do not interfere with each other. Another example is the *Out of Gas* vulnerability. On Ethereum, every transaction is associated with an upper bound on the amount of gas that can be spent, i.e., the amount of computation allowed. Any transaction will fail if the gas spent exceeds this limit. This vulnerability may be exploited in Denial-of-service (DoS) attacks, where the contract may become unavailable or unusable by being overwhelmed with requests or by consuming excessive amounts of resources.

**Prodigal smart contracts** are those that give away Ether to an arbitrary address that is not the owner or has never deposited Ether to the contract. Vulnerabilities arise that cause funds to be sent from the contract to a malicious user. An example is the *Unchecked Send* vulnerability, which happens due to lack of proper checking during transfer of ownership, paving the way for attackers to manipulate the send function.

**Suicidal smart contracts** are those where the source code contains the SUICIDE instruction and it can be triggered by a message sender that does not appear in the contract's state at the moment of receiving the message. With these contracts, the ownership can be seized and the contract may be killed, thereby locking the funds in the contract forever.

## 4. Transformer-Based Vulnerability Scanner (VASCOT)

Transformers have gained significant popularity in sequential analysis tasks [32]. Key components of transformers include *self-attention* mechanisms, which allow the model to weigh the parts of the input sequence and focus on specific parts during processing, and *positional encodings*, which inform the model about the position of each token in the sequence. Transformer-Based Vulnerability Scanner (VASCOT) utilizes the transformer architecture to detect the relations between the semantic representations in the smart contract opcode sequence.

Our model implementation is based on PyTorch. The input to our model consists of an embedding layer that converts the tokenized contract data into vector representations, with each token being represented by 64-dimensional vectors. The output of this layer is processed through a Transformer block which, unlike LSTM, allows our model to consider the entire sequence of data simultaneously, leveraging the self-attention mechanism.

The tokenization process breaks down the opcodes into tokens, which are then converted into numerical representations that can be processed by the model. Tokenized inputs are passed through an embedding layer, which transforms each token into a vector representation with a fixed dimension. This embedding process enables semantic information to be captured, allowing the model to represent similiar opcodes with vectors that are close in the embedding space.

VASCOT utilizes a self-attention mechanism, which allows the model to dynamically focus on different tokens in the sequence based on their relevance. Self-attention computes attention scores for each token pair, enabling the model to consider relationships between opcodes at various positions across the sequence. Thus, the model can weigh the importance of each token in understanding the overall structure within the contract. These attention scores are then used to produce weighted combinations of the embeddings, which act as the inputs of the following layers in the Transformer. Furthermore, positional encodings are added to the embeddings to guarantee that the model is aware of the layout of tokens in the sequence. These encodings help VASCOT differentiate between tokens based on their position within the sequence, preserving the information of order in the corresponding opcodes.

**Table 1**
Transformer Model Configuration

| Layer | Input Shape | Output Shape | Parameters |
|---|---|---|---|
| Embedding | (batch size, 2048) | (batch size, 2048, 128) | 128,000 |
| Transformer Encoder | (batch size, 2048, 128) | (batch size, 2048, 128) | 65,536 |
| Attention Pooling | (batch size, 2048, 128) | (batch size, 128) | 16,384 |
| Linear | (batch size, 128) | (batch size, 2) | 258 |

The output of the Transformer block then undergoes an Attention Pooling layer, which dynamically learns to focus on the most important parts of the sequence while reducing it to a 1D format while retaining crucial features.

We then apply two dropout layers with a rate of 0.2 to prevent overfitting. Between these, we include a dense layer with a ReLU activation function to add non-linearity and further refine learned features. Finally, we connect a dense layer with a sigmoid activation function as our output layer, which provides us the probability estimation for the vulnerability likelihood of each contract. The model configuration is shown in Table 1.

### 4.1. Sliding Window Implementation

The Transformer model has a limit on the sequence length it can handle in one pass; this is dominated by the *max_length* parameter. Our data set contains longer sequences than the *max_length* value that an accessible processor could handle. While we could address this issue by *truncating* the opcode sequences to *max_length*, this would potentially manipulate the data in a way where vulnerability traces did not exist in the feature set. Another alternative approach would be *chunking*, where the sequence is split into non-overlapping chunks smaller than the *max_length*; however, this approach could conceal important information if vulnerability features are scattered across separate chunks.

Since all of the input sequences is important in our dataset, we proceeded with another solution, and we process the input sequence in a *sliding window* approach. Any input sequence that is larger than *max_length* (e.g., 2048 tokens) is divided into overlapping chunks of tokens. These chunks are fed into the model sequentially, and the results from each chunk are aggregated to make predictions across the all input sequence. In our implementation, the overlap amount is 0.25 and the window size is 2048. The sliding window approach is demonstrated in Figure 2.

## 5. Data Sets

The experimental study in this work has been carried out on two separate datasets. We collected the smart contracts in the first dataset; the construction of the dataset is explained in detail in this section. The second dataset is a public dataset that contains more samples but dates back several years prior to the first dataset.

### 5.1. Dataset Comprising Verified Contracts

As Solidity language and solc compiler are enhanced through the years [33], the vulnerabilities present in smart contracts written/compiled in different versions of the language/compiler may change. To obtain a data set of recent contracts, we construct a labeled data set comprising the *verified* smart contracts deployed in 2022. Verified smart contracts have their source code (as well as exact compiler version and settings used) publicly available, such that a third-party platform such as Etherscan can recompile the provided source code and check if the resulting bytecode exactly matches the bytecode deployed on the Ethereum network. If the bytecodes match, the contract is marked as "verified".

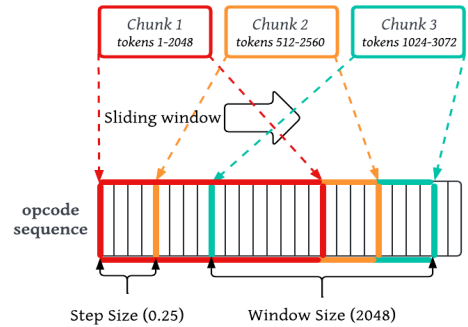

**Figure 2**: Sliding window implementation on VASCOT for handling contracts larger than 2,048 opcodes

#### 5.1.1. Accessing Smart Contracts

We used Etherscan [34] to access verified smart contracts. First, we created a developer key to interact with the Etherscan API via the Etherscan.Net.lib library. Using this API, we obtained a list of verified contract addresses that have open-source licenses. For each address in the .csv file, the API was invoked to retrieve the corresponding source code from the specified endpoint. Retrived contracts were saved using the ContractAddress_ContractName.sol naming convention, which helped to identify any duplicates.

While saving the contract source codes returned from the Etherscan API, we noticed cases where more than one .sol file was returned. In order to parse the source code correctly, we compared the .sol values in the *sourceCode* field of the JSON-formatted response. We extracted and saved only the contracts that contained a single ".sol" value.

The Etherscan API provides access to a random set of contracts typically ranging from 100 to 600 on a daily basis. Using this API, we were able to obtain the data of 16,363 *verified* contracts over an interval of four months between January 2022 and April 2022.

The JSON response returned over the Etherscan API also contains the bytecode, in addition to source code. To convert the EVM bytecode to EVM opcode hex, each byte in the bytecode is parsed into its respective hexadecimal representation, but without the leading 0*x*'s. For example, in the bytecode in Figure 1, the first opcode is PUSH1 0x80, which can be represented by the hexadecimal representation of 0x60 0x80. The next opcode is PUSH1 0x40, which is represented as 0x60 0x40. The respective opcodes for Figure 1 (without the leading 0x's) are 60 80 60 40 52 34 80 15 62 00 00 11 57 60 00 and so on.

The opcode sequence corresponding to each contract address is recorded, along with the corresponding label. This sequence constitutes the feature set for VASCOT, using which the vulnerability patterns will be identified.

### 5.1.2. Labeling the Data

To label the smart contracts, for each retrieved smart contract source code, we ran the analysis using MAIAN [35] to categorize the data into four classes representing smart contracts with no vulnerabilities, as well as those that demonstrate suicidal, greedy, and prodigal smart contract characteristics. In our trials, we encountered errors caused by MAIAN's incompatibility with the more recent smart contracts' compiler versions. For the collective solution to this problem in the new contract pool we create, we overwrote the compiler version to the range between Solidity v0.4.22 and v0.9.0 to work correctly with MAIAN.

MAIAN formulates the erroneous behaviours in terms of predicates of observed traces that are created during execution. For example, a transition label of the form *call(id, m)* captures the fact that a currently running contract has transferred control to another contract *id*, by sending it a message *m*. MAIAN checks if there is an execution trace that violates ETHERLITE rules. The predicates allow capturing repeating patterns in contract life cycle (*greedy*), the conditions indicating vulnerability to unauthorized access to a contract's funds (*prodigal*), or a contract's suicide functionality (*suicidal*).

In our output dictionary, the string "1 0 0 0" represents no vulnerabilities, "0 1 0 0" represents suicidal, "0 0 1 0" represents prodigal, and "0 0 0 1" represents greedy vulnerabilities. A .csv file containing the addresses of all smart contracts is formed; the final data set contains contract address, the opcode, and the respective label. A sample of contracts in our final data set are shown in Figure 3.

| SMART CONTRACT ADDRESS | OPCODE | LABEL |
|---|---|---|
| 0x000087bb453ff203eca1afb9a8d2b80fec94083b6 | 60 80 60 40 52 60 04 36 10 61 01 02 57 60 00 35 60 e0 1c 80 63 71 50 18 a6 11 61 00 | 0 0 0 1 |
| 0x0008ad3ea1bbda4b757d57b5c13a48fa7842d896 | 60 80 60 40 52 60 04 36 10 61 01 02 57 60 00 35 60 e0 1c 80 63 71 50 18 a6 11 61 00 | 0 0 0 1 |
| 0x000c766455346ea24fd8575ef74429b90740a834 | 73 00 0c 76 64 55 34 6e a2 4f d8 57 5e f7 44 29 b9 07 40 a8 34 30 14 60 80 60 40 52 6 | 1 0 0 0 |
| 0x002e79d50e61a73de91edf3c82a8f2577f9e680d | 60 80 60 40 52 34 80 15 61 00 10 57 60 00 80 fd 5b 50 60 04 36 10 61 00 cf 57 60 00 | 1 0 0 0 |
| 0x006829f48dc2448a3d306b5890ff14f4572e029c | 73 00 68 29 f4 8d c2 44 8a 3d 30 6b 58 90 ff 14 f4 57 2e 02 9c 30 14 60 80 60 40 52 60 | 1 0 0 0 |
| 0x00897c8ec7af97aeebe1a46bb543035436184485 | 60 80 60 40 52 60 04 36 10 61 01 23 57 60 00 35 60 e0 1c 80 63 75 10 39 fc 11 61 00 | 0 0 0 1 |
| 0x00984d4c5445476c7c0183bf27ef2f94e0194698 | 60 80 60 40 52 60 04 36 10 61 01 39 57 60 00 35 60 e0 1c 80 63 6f c3 ea ec 11 61 00 a | 0 0 0 1 |
| 0x009ab9bf0c32151531e9896d7737a6b7acc81aed | 60 80 60 40 52 34 80 15 61 00 10 57 60 00 80 fd 5b 50 60 04 36 10 61 00 cf 57 60 00 | 1 0 0 0 |
| 0x00c06dc31daa37937a911b8c35ea85ea3f1bf044 | 60 80 60 40 52 60 04 36 10 61 01 2e 57 60 00 35 60 e0 1c 80 63 6d dd 17 13 11 61 00 | 1 0 0 0 |

**Figure 3**: Part of the final data set.

### 5.1.3. Preprocessing

The Etherscan API returns a random set of contracts upon each query, and our observations revealed many duplicates among the collected contracts in the data set. After removing the duplicates, the contracts that were flagged as vulnerable contained 8,559 contracts with the suicidal flag set to 1 and 7,803 contracts with greedy flag set to 1. We observed that our data set contained only 1 contract with the prodigal flag set to 1. We attribute this to the fact that the obtained data set comprises only contracts that have been verified via an audit process, and that the prodigal behavior is explicit and therefore can be caught and prevented. The single prodigal contract, as well as the duplicate contracts that were reported to be both suicidal and prodigal, were removed from the data set. After removing duplicates and prodigal contracts, 7,231 unique contracts remained.

Additionally, some contracts were flagged to be both greedy and suicidal. For an accurate classification into the sub-classes, we wanted to eliminate any obfuscation; thus, we removed such entries from the data set. After removing such entries, our dataset contained contracts that were each flagged in a single vulnerability category.

MAIAN facilitates symbolic execution of the contract bytecode on a private fork of Ethereum blockchain using concrete transaction values, to validate the contract exhibits vulnerable behavior in concrete execution. Contracts that do not exhibit specified traces are identified as false positives. Upon concrete validation on a private fork of Ethereum, it was observed that the greedy and suicidal contracts in our dataset contained false positives. When false positives were also removed, **1,915** total unique contracts remained in the data set, containing **1,202** non-vulnerable contracts and **713** greedy contracts.

Figure 4 demontrates the distribution of contract length (number of opcodes) for all verified smart contracts in our data set. We observe that the length of contracts in these figures, particularly those that are flagged as vulnerable, are longer (average length of ∼6000) compared to the distribution of Google BigQuery data set explained in Sec. 5.2, where the average contract length is ∼1500 opcodes. Verified contracts that are active in Ethereum as of 2022 have longer opcode lengths.

### 5.2. Dataset Comprising Unverified Contracts

In order to measure the generalizability of VASCOT and compare it with that of LSTM, we also used a large public data set, the Google BigQuery data set [36]. This dataset contains a total of 891,162 contracts, including all contracts from the first block of Ethereum, up until block 4,799,998, which was the last block mined on December 26, 2017. It has been used in previous studies [8],[9].

In [9], this dataset was preprocessed, and 8,469 distinct contracts were identified as vulnerable, with **1,207** of them flagged as suicidal, **1,461** of them prodigal, and **5,801** of them greedy. We used these vulnerable contracts in our experiments. After removing invalid opcode instructions or duplicates, the public dataset contained 416,944 unflagged contracts. From this set, we selected a subset of 8,000 unflagged contracts and thus constructed a balanced data set comprising 16,469 contracts.

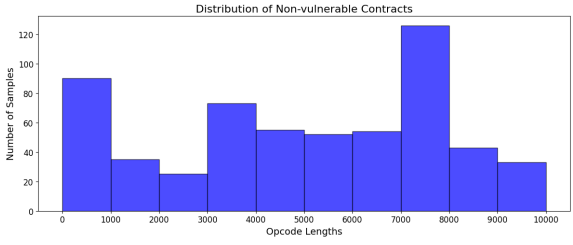

(a) Non-vulnerable contracts

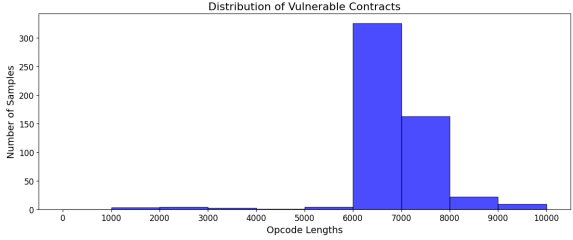

(b) Vulnerable contracts

**Figure 4:** Distribution of the number of vulnerable and non-vulnerable smart contracts in our collected data set.

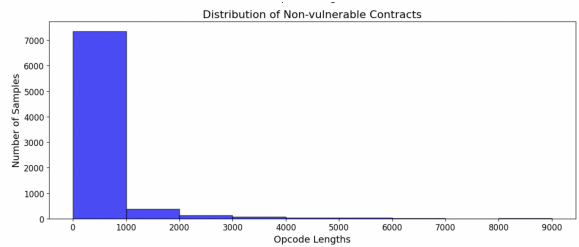

(a) Non-vulnerable contracts

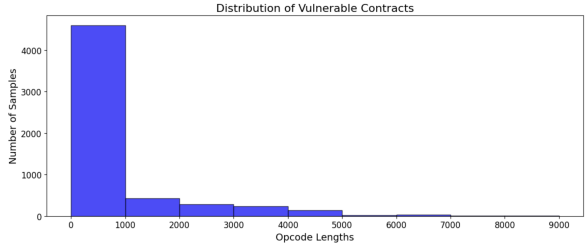

(b) Vulnerable contracts

**Figure 5:** Distribution of the number of vulnerable and non-vulnerable smart contracts in the Unverified data set containing 16,469 contracts.

Figure 5 demonstrates the histograms of the length of smart contracts in terms of the number of opcodes in the compiled opcode, for vulnerable and non-vulnerable contracts. On these figures we have two main observations; (i) the vulnerable contracts are slightly longer in length, and (ii) the smart contracts in this data set have at most 7000 opcodes and most of the smart contracts contain less than 1500 opcodes.

Next, we also generated a data set comprising only the type of vulnerabilities that exist in the 2022 dataset. Thus, we included only the **5,801** greedy vulnerable contracts and

(a) Non-vulnerable contracts

(b) Vulnerable contracts

**Figure 6:** Distribution of the number of greedy and non-vulnerable smart contracts in the Unverified data set containing 13,801 contracts.

**Table 2**
Smart Contract Data Sets

|  | Date | Normal Count | Vulnerable Count | Vuln. | Trans-parency |
|---|---|---|---|---|---|
| Dataset I | 2022 | 1,202 | 713 | G | Verified |
| Dataset II | 2017 | 8,000 | 8,469 | S,P,G | Unverified |
| Dataset II-r | 2017 | 8,000 | 5,801 | G | Unverified |

**8,000** non-vulnerable contracts and combined them into a dataset of size 13,801. We plot the distribution of length for vulnerable and non-vulnerable smart contracts in Figure 6. The distribution of compiled contract length for vulnerable and non-vulnerable contracts for this data set closely match those in the complete unverified data set (containing all three vulnerability types) but strictly differs from the respective distributions in the newly constructed 2022 data set.

## 6. Experiments

In our experiments, we evaluate the performance of VASCOT on the different smart contract data sets explained in Sec. 5, and compare it with the LSTM model in terms of accuracy, precision and recall for identifying vulnerabilities. The datasets are summarized in Table 2. In the following, we first provide model implementation details, then explain each setup and our observations therein.

### 6.1. Implementation Details

The experiments are carried out on NVIDIA TESLA P100 GPU.

The dataset was divided into train, validation, and test partitions at 68%, 15%, and 18% proportions, respectively. We study the both the binary classification problem of identifying vulnerable contracts, as well as multi-class classification of individual vulnerability types.

Table 3
Optimized Model Parameters

| Parameter | VASCOT | LSTM |
|---|---|---|
| max_length | 2048 | 2048 |
| embedding_dim | 128 | 128 |
| batch_size | 32 | 32 |
| learning_rate | 0.0005 | 0.0005 |
| head_size | 128 | - |
| head_dim | - | 256 |
| dropout | 0.2 | 0.2 |
| num_heads | 4 | - |
| ff_dim | 4*128 | - |

Transformer and LSTM common parameters are shown in Table 3. In our previous VASCOT implementation in [37] we had set the maximum input length (*max_length*) to 1024, because the server could not handle higher values, and we had zero-padded the contracts that were shorter. In this work, we run the experiments on a stronger server, and set this value to 2048, and the sliding window mechanism is activated for input sequences that are larger than this value.

Number of distinct input tokens, i.e., *num_words* for the Transformer is dynamically varied in the experiments based on the vocabulary that is constructed by the tokenizer. While the tokenizer found 150 tokens in some sequences, other sequences with 270 distinct words were also detected. It must be noted that although there are 140 distinct opcodes in EVM [26], the input sequences also include stack addresses or constant values.

We used Adam optimizer and binary cross-entropy loss function in both LSTM and Transformer models. The models converged in different epochs in the different scenarios tested; these values are specified under individual experiments.

### 6.2. Experiments on Dataset I: Verified Contracts

First, we comparatively study the performances of LSTM and VASCOT on Dataset I. This dataset contains only greedy vulnerabilities and non-vulnerable contracts.

Models were trained for 50 epochs. The training accuracy for LSTM and VASCOT models is 99.6% and 95.2%, respectively, while the validation accuracy values are 90.6% and 96.2% for these models. We observe that LSTM slightly overfits while VASCOT learns the relations in opcode sequences much better and can generalize better than LSTM. The training and validation loss values for LSTM strengthen this observation; while LSTM has a training loss of 0.014, the validation loss is 0.598. For VASCOT these values are 0.136 and 0.125, respectively.

Table 4 shows the test performance of both models on this data set, where **L** is for LSTM and **V** is for VASCOT. Testing has been performed on 326 samples, LSTM obtained an accuracy of 88% and VASCOT's accuracy was 95%. VASCOT outperformed LSTM in detecting both non-vulnerable and greedy contracts, despite the small training set size. As the problem at hand aims to detect vulnerability, it is also valuable that VASCOT attains 15% higher precision and 7% higher recall than LSTM in detecting vulnerable samples (class "1").

Table 4
Binary Classification Performance on Verified Contracts

|   | Precision | | Recall | | f1-Score | | Support |
|---|---|---|---|---|---|---|---|
|   | L | V | L | V | L | V | |
| 0 | 0.90 | 0.94 | 0.92 | 0.99 | 0.91 | 0.96 | 205 |
| 1 | 0.85 | 0.98 | 0.83 | 0.89 | 0.84 | 0.94 | 121 |

Figure 7 shows the confusion matrices of both models when trained and tested on verified contracts. VASCOT almost manages to output no false positives, and provides significant reduction in false negatives compared to LSTM model.

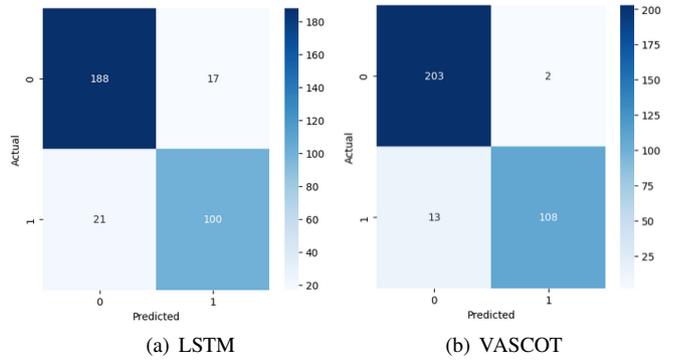

(a) LSTM  (b) VASCOT

**Figure 7:** Confusion matrices for LSTM vs VASCOT on verified contracts.

### 6.3. Experiments on Dataset II: Unverified Contracts

Next, the two models are compared (*i.e.,* trained and tested) on Dataset-II. This set comprises all three types of vulnerabilities as shown in Table 2. We evaluate the model's success in identifying any type of vulnerable contracts.

Training converged much faster on this dataset; specifically, while training for 50 epochs was needed on Dataset I, 20 epochs were sufficient on Dataset II. Similar to what was observed in Dataset I, the LSTM model had a training accuracy of 99.3% and a validation accuracy of 96.3%; VASCOT had a training accuracy of 93.3% and a validation accuracy of 94.2%. LSTM had a training loss of 0.022 and a validation loss of 0.164 indicating overfitting; VASCOT had a training loss of 0.179 and validation loss of 0.169 demonstrating it can better generalize to unseen sequences.

Table 5 shows the test performance of both models on this data set, where **L** is for LSTM and **V** is for VASCOT, and **S**, **P**, **G** correspond to suicidal, prodigal and greedy vulnerability classes. Although VASCOT and LSTM demonstrate close performance for non-vulnerable contracts, we observe that LSTM outperforms VASCOT in all vulnerability classes; we believe this is due to overfitting on the small number of samples. This result is in support of the previous work [9], where the reported accuracy was primarily due to correctly detecting the large number of (specifically, 200,000) non-vulnerable samples.

**Table 5**
Multi-Class Performance in Dataset-II

|   | Precision | | Recall | | f1-Score | | Support |
|---|---|---|---|---|---|---|---|
|   | L | V | L | V | L | V |   |
| 0 | 0.98 | 0.97 | 0.98 | 0.96 | 0.98 | 0.96 | 1362 |
| S | 0.94 | 0.92 | 0.96 | 0.85 | 0.95 | 0.88 | 205 |
| P | 0.88 | 0.79 | 0.87 | 0.81 | 0.87 | 0.80 | 248 |
| G | 0.96 | 0.92 | 0.95 | 0.94 | 0.96 | 0.93 | 983 |

Figure 8 shows the confusion matrices of both models in this scenario. We observe that among all classes, VASCOT performs poorest in detecting prodigal and suicidal vulnerabilities. This lower performance can be insignificant in assessing the applicability and contribution of VASCOT today, considering the suicidal and prodigal behaviors have significantly been eliminated in recent contracts (hence they are unseen in Dataset I), both due to improvements in Solidity language and the use of auditing practices prior to deployment. It must also be highlighted that the contracts in this dataset were *unverified*, suggesting this dataset may not be suitable for benchmarking.

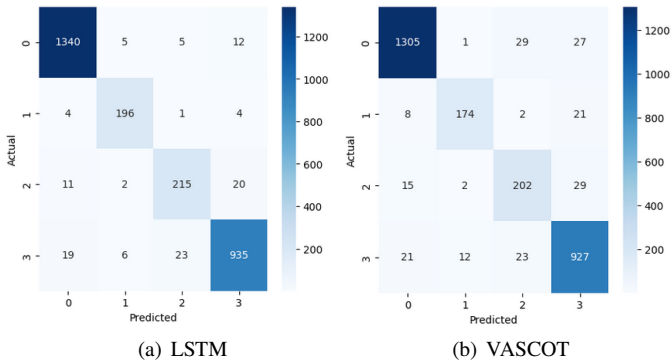

(a) LSTM  (b) VASCOT

**Figure 8:** Confusion matrices for LSTM and VASCOT on Dataset II.

### 6.4. Experiments on Dataset II-r: Unverified Filtered Contracts

Dataset II contains vulnerability classes that have not been found in the recent contracts. To provide a fair comparison, we removed the suicidal and prodigal contracts and constructed Dataset II-r (representing reduced Dataset II) comprising 5,801 greedy contracts and 8,000 non-vulnerable contracts.

On this dataset, LSTM had a training accuracy of 99.73% and validation accuracy of 98.21%; the respective loss values were 0.009 and 0.078, demonstrating 7.6 times increase. VASCOT yielded training and test accuracies of 96.86% and 97.10%, respectively, and loss values were 0.081 and 0.077. Similar to the results observed on Dataset II, VASCOT demonstrates better generalizability and LSTM may be overfitting. The test accuracy of the models were 0.98 for LSTM and 0.97 for VASCOT.

**Table 6**
Binary Classification Performance in Dataset II-r

|   | Precision | | Recall | | f1-Score | | Support |
|---|---|---|---|---|---|---|---|
|   | L | V | L | V | L | V |   |
| 0 | 0.98 | 0.97 | 0.98 | 0.97 | 0.98 | 0.97 | 1361 |
| 1 | 0.97 | 0.96 | 0.97 | 0.96 | 0.97 | 0.96 | 983 |

**Table 7**
Binary Classification Performance in Unverified-Verified Cross-Dataset Evaluation (L: LSTM, T: Transformer)

|   | Precision | | Recall | | f1-Score | | Support |
|---|---|---|---|---|---|---|---|
|   | L | V | L | V | L | V |   |
| 0 | 0.73 | 0.74 | 0.27 | 0.67 | 0.39 | 0.70 | 1202 |
| 1 | 0.40 | 0.52 | 0.83 | 0.60 | 0.54 | 0.55 | 713 |

Table 6 demonstrates the precision, recall, f-score of both models on Dataset II-r. We observe very close performance from the two models under consideration.

### 6.5. Cross-Dataset Analysis

To examine the generalizability of the two models, we also perform cross-dataset analysis. Specifically, we train on unverified contracts and test on verified contracts, and vice versa. In these experiments, the Dataset II-r is used for the unverified contracts, as we would like the model to only be trained and tested to recognize the classes that are present in both datasets.

#### 6.5.1. Train on Unverified, Test on Verified

First, we train the model on all of the contracts in Dataset II-r, and test it on all the contracts in Dataset I. Training was performed in 20 epochs.

Training and validation accuracy for LSTM was 99.78% and 98.36%, respectively; the respective loss values were 0.007 and 0.079. Using VASCOT, training and validation accuracy values were 97.30% and 97.73%, and the loss values were 0.072 and 0.068, respectively. The difference in the training and validation performances with LSTM points out a potential overfitting problem.

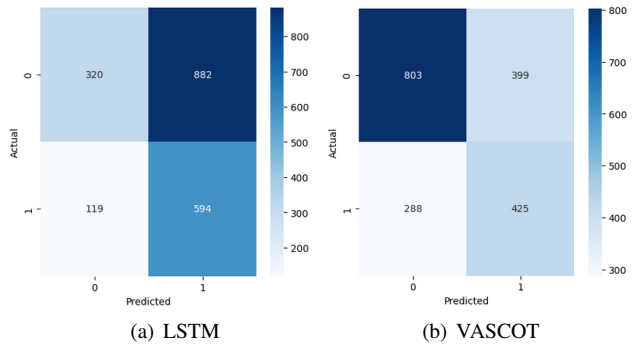

(a) LSTM  (b) VASCOT

**Figure 9:** Confusion matrices for the unverified-verified cross-dataset analysis.

The test accuracy of LSTM on 1915 samples was 48% and that of VASCOT was 64%. Additionally, Table 7 demonstrates higher precision using VASCOT for both normal and vulnerable classes. We observe that LSTM model cannot generalize to detect the non-vulnerable contracts in the recent dataset. It may be due to the average contract length being larger in the recent contracts, and the Transformer models generally handling long-term dependencies better than LSTM, thanks to the self-attention mechanism that allows them to weigh relationships between *all* tokens in a sequence regardless of their distance. A comparison of figures 4(a) and 6(a) demonstrates that the length of non-vulnerable contracts differs between the two datasets. The difference is even more highlighted for the vulnerable contracts as observed in figures 4(b) and 6(b). It is also expected that the sequences differ significantly; some older vulnerabilities are no longer observed in the recent contracts as developers become more informed about the potential threat imposed by certain patterns [38]. It must be noted that VASCOT outputs many false negatives in this scenario (Figure 9), which necessitates a deeper understanding of the attack patterns; we leave this investigation of explaining the model behavior to our future study.

**Table 8**
Verified-Unverified Cross-Dataset Evaluation

|   | Precision | | Recall | | f1-Score | | Support |
|---|---|---|---|---|---|---|---|
|   | L | V | L | V | L | V |   |
| 0 | 0.62 | 0.62 | 0.99 | 0.95 | 0.76 | 0.75 | 8006 |
| 1 | 0.94 | 0.74 | 0.15 | 0.21 | 0.25 | 0.33 | 5782 |

### 6.5.2. Train on Verified, Test on Unverified

Our next experiment investigates the performance of the models when trained on the verified contracts and tested using the unverified samples. As there were only 1,915 contracts in the training set, the model training took 75 epochs to converge.

Similar to the prior observations, LSTM training and validation accuracy and loss values indicated potential overfitting; 99.94% and 89.24% respective accuracies and 0.005 and 0.681 loss values were attained in training and validation, respectively. The training and validation performance with VASCOT did not vary between training and validation; observed accuracy values were 94.84% and 95.49%, and loss values were 0.153 and 0.122 for training and validation, respectively. In all experiments we consistently observe that the Transformer model can better generalize new sequences.

The test accuracy of both models is 64%, and as shown in Table 8. Even though the models can detect non-vulnerable contracts, they are challenged in detecting vulnerable ones. While VASCOT performs better compared to LSTM in this scenario (Figure 10), it still outputs many false negatives. This scenario offers both the challenge of cross-dataset training and testing, and the fact that the training set is very small (Table 2). In summary, both cross-dataset experiments demonstrated challenges in the generalizability of the two models. The two datasets were collected four years apart, and they significantly differed as seen from the contract lengths in Figures 4 and 5. In our next work, we will apply transfer learning to better adapt the models on the new datasets and discuss our observations.

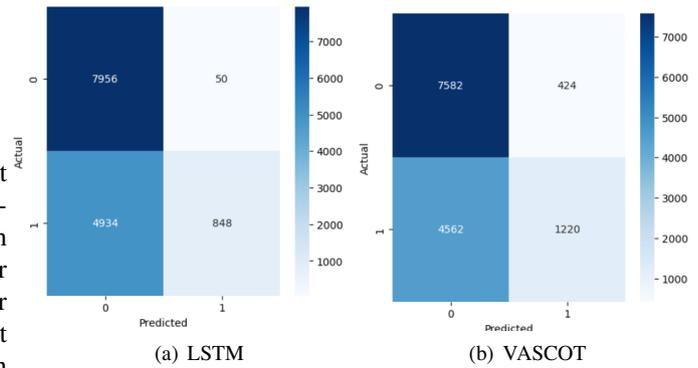

(a) LSTM    (b) VASCOT

**Figure 10:** Confusion matrices for the verified-unverified cross-dataset analysis.

## 7. Conclusions and Future Work

Smart contracts can be used for various scenarios, from automating financial transactions in decentralized finance (DeFi) to managing supply chain logistics, or ensuring the integrity of digital assets such as non-fungible tokens (NFTs). Due to the large amounts of money that are transferred through smart contracts, these programs often become the targets of malicious actors attempting to exploit their vulnerabilities. These vulnerabilities involve decision logic, branching, sequencing, and interaction with other Ethereum addresses, and therefore are challenging to detect. In this work, we studied vulnerability detection in smart contracts using deep learning tools that sequentially scan the contract bytecode/opcode.

We proposed a smart contract vulnerability scanner, VASCOT, that utilizes the Transformer model. We then constructed a data set containing the opcodes of 16,469 verified Ethereum smart contracts. We evaluated the success of VASCOT in comparison with a prior sequential analysis tool using LSTM, on the constructed data set, as well as on an older public data set, the choice of which was driven from its prior adoption by literature. Our results showed that VASCOT outperforms LSTM in terms of generalizability to unseen sequences; the most significant benefit being observed on the recent dataset comprising only verified contracts that were longer in compiled length.

Our future work will include examining the contribution of specific sequences on the models to explain the model behavior. Additionally, we will enhance our cross-dataset analysis by implementing fine tuning and transfer learning of the models and comparatively analyzing their performances.

## References

[1] Wikipedia, The Free Encyclopedia, The DAO, https://en.wikipedia.org/w/index.php?title=The_DAO&oldid=1240033956, 2024. [Online].
[2] S. Palladino, The parity wallet hack explained, https://blog.openzeppelin.com/on-the-parity-wallet-multisig-hack-405a8c12e8f7, 2017.

[3] ImmuneBytes, bzx protocol exploit – detailed analysis, https://www.immunebytes.com/blog/bzx-protocol-exploit-sep-14-2020-detailed-analysis/, 2022.

[4] T. Pearson, Top 10 crypto hacks of 2023, https://www.dlnews.com/articles/defi/top-10-crypto-hacks-of-2023-ranked-as-stakecom-is-fifth/, 2023.

[5] S. T. et al., Smartcheck: Static analysis of ethereum smart contracts, in: IEEE/ACM 1st International Workshop on Emerging Trends in Software Engineering for Blockchain (WETSEB), 2018, pp. 9–16.

[6] S. Akca, A. Rajan, C. Peng, Solanalyser: A framework for analysing and testing smart contracts, in: IEEE 26th Asia-Pacific Software Engineering Conference (APSEC), 2019, pp. 482–489.

[7] L. Luu, D.-H. Chu, H. Olickel, P. Saxena, A. Hobor, Making smart contracts smarter, in: Proceedings of the 2016 ACM SIGSAC Conference on Computer and Communications Security (CCS), 2016, p. 254–269.

[8] I. Nikolić, A. Kolluri, I. Sergey, P. Saxena, A. Hobor, Finding the greedy, prodigal, and suicidal contracts at scale, in: Proceedings of the 34th Annual Computer Security Applications Conference, 2018, p. 653–663.

[9] W. J.-W. Tann, X. J. Han, S. S. Gupta, Y.-S. Ong, Towards safer smart contracts: A sequence learning approach to detecting security threats, arXiv preprint arXiv:1811.06632 (2018).

[10] Y. Zhang, D. Liu, Toward vulnerability detection for ethereum smart contracts using graph-matching network, Future Internet 14 (2022) 326.

[11] N. F. Samreen, M. H. Alalfi, Reentrancy vulnerability identification in ethereum smart contracts, in: 2020 IEEE International Workshop on Blockchain Oriented Software Engineering (IWBOSE), 2020, pp. 22–29.

[12] Q. Zhao, C. Huang, L. Dai, Vuldeff: Vulnerability detection method based on function fingerprints and code differences, Knowledge-Based Systems 260 (2023) 110139.

[13] A framework for bug hunting on the ethereum blockchain, 2017. URL: https://github.com/ConsenSys/mythril.

[14] M. Mossberg, F. Manzano, E. Hennenfent, A. Groce, G. Grieco, J. Feist, T. Brunson, A. Dinaburg, Manticore: A user-friendly symbolic execution framework for binaries and smart contracts, in: 2019 34th IEEE/ACM International Conference on Automated Software Engineering (ASE), 2019, pp. 1186–1189.

[15] S. Kalra, S. Goel, M. Dhawan, S. Sharma, Zeus: analyzing safety of smart contracts., in: Ndss, 2018, pp. 1–12.

[16] P. Tsankov, A. Dan, D. Drachsler-Cohen, A. Gervais, F. Buenzli, M. Vechev, Securify: Practical security analysis of smart contracts, in: Proceedings of the 2018 ACM SIGSAC conference on computer and communications security, 2018, pp. 67–82.

[17] P. Qian, Z. Liu, Q. He, R. Zimmermann, X. Wang, Towards automated reentrancy detection for smart contracts based on sequential models, IEEE Access 8 (2020) 19685–19695.

[18] A. K. Gogineni, S. Swayamjyoti, D. Sahoo, K. K. Sahu, R. Kishore, Multi-class classification of vulnerabilities in smart contracts using AWD-LSTM, with pre-trained encoder inspired from natural language processing, IOP SciNotes 1 (2020) 035002.

[19] W. Wang, J. Song, G. Xu, Y. Li, H. Wang, C. Su, Contractward: Automated vulnerability detection models for ethereum smart contracts, IEEE Transactions on Network Science and Engineering 8 (2021) 1133–1144.

[20] L. Zhang, J. Wang, W. Wang, Z. Jin, Y. Su, H. Chen, Smart contract vulnerability detection combined with multi-objective detection, Computer Networks 217 (2022) 109289.

[21] X. Ren, Y. Wu, J. Li, D. Hao, M. Alam, Smart contract vulnerability detection based on a semantic code structure and a self-designed neural network, Computers and Electrical Engineering 109 (2023) 108766.

[22] C. Sendner, H. Chen, H. Fereidooni, L. Petzi, J. König, J. Stang, A. Dmitrienko, A.-R. Sadeghi, F. Koushanfar, Smarter contracts: Detecting vulnerabilities in smart contracts with deep transfer learning, in: NDSS, 2023.

[23] D. Chen, L. Feng, Y. Fan, S. Shang, Z. Wei, Smart contract vulnerability detection based on semantic graph and residual graph convolutional networks with edge attention, Journal of Systems and Software 202 (2023) 111705.

[24] X. Sun, L. Tu, J. Zhang, J. Cai, B. Li, Y. Wang, Assbert: Active and semi-supervised bert for smart contract vulnerability detection, Journal of Information Security and Applications 73 (2023) 103423.

[25] S. Bistarelli, G. Mazzante, M. Micheletti, L. Mostarda, F. Tiezzi, Analysis of ethereum smart contracts and opcodes, in: Advanced Information Networking and Applications: Proceedings of the 33rd International Conference on Advanced Information Networking and Applications (AINA-2019) 33, Springer, 2020, pp. 546–558.

[26] G. Wood, et al., Ethereum: A secure decentralised generalised transaction ledger, Ethereum project yellow paper 151 (2014) 1–32.

[27] Ethereum virtual machine opcodes, 2017. URL: https://www.ethervm.io/.

[28] D. Muhs, Smart Contract Weakness Classification (SWC) Registry, 2020. URL: https://github.com/SmartContractSecurity/SWC-registry.

[29] P. Zhang, F. Xiao, X. Luo, A framework and dataset for bugs in ethereum smart contracts, in: 2020 IEEE international conference on software maintenance and evolution (ICSME), 2020, pp. 139–150.

[30] IEEE Standard Classification for Software Anomalies, IEEE Std 1044-2009 (Revision of IEEE Std 1044-1993) (2010) 1–23.

[31] I. Sergey, A. Hobor, A concurrent perspective on smart contracts, in: Financial Cryptography and Data Security: FC 2017 International Workshops, Springer, 2017, pp. 478–493.

[32] Q. Wen, T. Zhou, C. Zhang, W. Chen, Z. Ma, J. Yan, L. Sun, Transformers in time series: A survey, 2023. arXiv:2202.07125.

[33] A. Pinna, S. Ibba, G. Baralla, R. Tonelli, M. Marchesi, A massive analysis of ethereum smart contracts empirical study and code metrics, IEEE Access 7 (2019) 78194–78213.

[34] Etherscan Export CSV Data - List of Verified Contract addresses with an open source license, 2024. URL: https://etherscan.io/exportData?type=open-source-contract-codes.

[35] A. Gritz, MaianUpdated, 2022. URL: https://github.com/alegritz/MAIANUpdated/tree/main/MAIAN-master.

[36] Google BigQuery, 2018. URL: https://cloud.google.com/blog/products/data-analytics/ethereum-bigquery-public-dataset-smart-contract-analytics.

[37] E. Balcı, G. Yılmaz, A. Uzunoğlu, E. G. Soyak, Accelerating smart contract vulnerability scan using transformers, in: 2023 IEEE Asia-Pacific Conference on Computer Science and Data Engineering (CSDE), 2023, pp. 1–6.

[38] Y. Huang, Y. Bian, R. Li, J. L. Zhao, P. Shi, Smart contract security: A software lifecycle perspective, IEEE Access 7 (2019) 150184–150202.